# Cyber Campaign Fractals

*Geometric Analysis of Hierarchical Cyber Attack Taxonomies*

Ronan Mouchoux, University of South Brittany
François Moerman, XRATOR

**Abstract.** This paper introduces a novel mathematical framework for analyzing cyber threat campaigns through fractal geometry. By conceptualizing hierarchical taxonomies (MITRE ATT&CK, DISARM) as snowflake-like structures with tactics, techniques, and sub-techniques forming concentric layers, we establish a rigorous method for campaign comparison using Hutchinson's Theorem and Hausdorff distance metrics. Evaluation results confirm that our fractal representation preserves hierarchical integrity while providing a dimensionality-based complexity assessment that correlates with campaign complexity. The proposed methodology bridges security taxonomy analysis and computational geometry, providing analysts with both mathematical rigor and interpretable visualizations for addressing the growing complexity of adversarial operations across multiple threat domains.

**Keywords:** Cybersecurity, Fractal Analysis, Threat Modeling, MITRE ATT&CK, DISARM Framework, Threat Intelligence, Similarity Metrics

*"Fractal geometry is not just a chapter of mathematics, but one that helps everyman to see the same world differently."*
The Fractal Geometry of Nature (1982), Benoît Mandelbrot

# 1 Introduction

The systematic analysis of adversarial campaigns represents one of the most pressing challenges in contemporary security research. Whether in the cyber domain (Advanced Persistent Threat groups, APT) or information operations (disinformation campaigns), security analysts face increasingly complex, multi-faceted threats that deploy diverse tactics, techniques, and procedures. Current approaches rely heavily on manual classification using taxonomic frameworks such as MITRE ATT&CK for cyber threats and DISARM for disinformation operations [1][2].

These taxonomies organize adversarial behaviors into hierarchical categories, typically using a three-level structure organized around a focal point:

- Campaign: a grouping of adversaries' behaviors with a specific timeframe and targeting.
- Tactics: High-level adversarial goals (e.g., "Initial Access" or "Establish Legitimacy").
- Techniques: Methods to achieve those goals (e.g., "Phishing" or "Create Fake Experts").
- Sub-techniques: Specific implementations (e.g., "Spearphishing Attachment" or "Use Fake Academic Credentials").

While taxonomically robust, these frameworks lack formal mathematical representations that would enable rigorous comparison, aggregation, and detection of structural relationships between campaigns.

We propose a fundamental reconceptualization: viewing hierarchical taxonomy structures as geometric objects with fractal properties. This insight emerged from visualizing campaign components as concentric layers (tactics at the center, techniques in the middle layer, sub-techniques in the outer layer), revealing structures reminiscent of snowflakes when projected on a plane. This visual analogy introduces a powerful mathematical framework—fractal geometry—that aligns naturally with the hierarchical, self-similar organization of adversarial tactics.

This paper makes the following contributions:
1. Formally map security taxonomies and fractal geometric structures.
2. A mathematical framework based on Iterated Function Systems (IFS), Hutchinson's Theorem, and Hausdorff distance metrics.
3. An algorithm for comparing campaign structures
4. Validation through case studies across multiple security domains
5. A dimensionality-based complexity measure

A crucial insight of our approach is its domain-agnostic nature. This enables comparative analysis across traditionally separate security domains (cyber operations, information warfare, physical security) using a unified mathematical language.

# 2 Related Works

## 2.1 Threat Campaign Modeling

Existing approaches to threat campaign analysis have primarily focused on graph-based representations. In the cyber domain, attack graphs [3] and kill chains [4] provide process-oriented views of attacker progression. For information operations, network representations of narratives [5] and communication patterns [6] seem to dominate literature.

The MITRE ATT&CK framework has become the de facto standard for cyber threat analysis [7]. DISARM frameworks provide structured taxonomies for information operations [1]. However, these frameworks lack formal mathematical representations for comparative analysis.

Recent work by Lin et al. introduced a systematic approach to labeling attack tactics specifically tailored for cyber threat hunting [8]. Their methodology emphasizes the importance of structured taxonomic labeling in facilitating effective threat detection and response, reinforcing the need for rigorous analytical frameworks capable of capturing hierarchical tactic structures.

## 2.2 Fractals in Complex Systems Analysis

Fractal geometry has been successfully applied to various complex systems, including financial networks [9], urban growth patterns [10], and information propagation in social networks [11]. Faloutsos et al. demonstrated that internet topology exhibits self-similar patterns consistent with fractal structures [12].

In cybersecurity research specifically, fractal dimension has emerged as a promising metric for anomaly detection and threat attribution. For example, Siddiqui and Khan applied fractal analysis to APT's network traffic for anomaly detection [13]. They also used fractal dimensions to characterize polymorphic malware behavior effectively [14]. Their results demonstrated that fractal-based features significantly improve classification accuracy compared to traditional methods, highlighting the potential of fractal geometry in cybersecurity analytics. However, the application of fractal geometry to taxonomy-based threat campaign behavioral structure remains unexplored.

## 2.3 Machine Learning Techniques in Threat Attribution and Detection

Machine learning techniques have increasingly been employed for advanced persistent threat (APT) attribution and detection due to their ability to handle complex data patterns effectively. Charan et al. systematically reviews various machine learning approaches and Data Mining techniques applied to APT scenarios (DMAPT) [15]. It highlights the strengths and limitations of existing methods, emphasizing the need for robust feature extraction techniques capable of capturing structural characteristics inherent in sophisticated threats.

Similarly, Mohamed and Belaton proposes a behavior-based detection approach utilizing machine learning classifiers trained on anomalous credential usage patterns in APT attacks (Strange Behaviour Inspection, SBI) [16]. This work underscores the value of behavior-driven features in identifying subtle indicators of compromise within hierarchical attack structures.

## 2.4 Structural Comparison Methods

Graph similarity metrics seem to dominate structural comparison approaches, with recent work by Koutra et al. on fast approximate graph alignment showing promise for large-scale analysis [17]. The application of Hausdorff distances to structured data comparison has precedents in bioinformatics [18] and pattern recognition [19].

## 2.5 Positioning

Our proposed geometric framework uniquely integrates fractal geometry with hierarchical security taxonomies, addressing gaps identified across existing literature:

- Providing a mathematical and intuitive method grounded in fractal geometry.
- Unlike prior fractal-based cybersecurity studies that primarily focus on traffic or malware analysis, we apply fractals directly to hierarchical taxonomy modeling.
- Our approach complements existing machine learning-driven methods by offering robust geometric feature extraction that can enhance predictive modeling accuracy.
- By formalizing hierarchical taxonomies into geometric objects, our method provides a unified mathematical language suitable for cross-domain security analysis—an aspect not addressed explicitly by prior literature.

Thus, our paper bridges these previously separate research streams—fractal geometry applications, hierarchical taxonomy modeling, structural comparison metrics, and machine learning approaches—into a cohesive analytical framework applicable across cybersecurity domains.

# 3 Theoretical Framework

## 3.1 Hierarchical Taxonomy Formalization

We begin by formalizing hierarchical security taxonomies as multi-level ordered sets. Let a campaign $C$ consist of a collection of tactics $T_i$, each containing techniques $\tau_{ij}$, which may further contain sub-techniques $\sigma_{ijk}$:

$$C = \{T_i\}_{i=1}^n, \quad T_i = \{\tau_{ij}\}_{j=1}^{m_i}, \quad \tau_{ij} = \{\sigma_{ijk}\}_{k=1}^{p_{ij}}$$

This formalization applies equally to cyber campaigns (using MITRE ATT&CK) and information operations (using DISARM).

## 3.2 Fractal Representation via Iterated Function Systems

We map this hierarchical structure to fractal geometry through an Iterated Function System (IFS), where each taxonomic element corresponds to a contraction mapping in $\mathbb{R}^2$:

$$\mathcal{F}_C = \{f_1, f_2, \ldots, f_n\}$$

where each $f_i$ is a contraction with factor $r_i < 1$. For hierarchical taxonomies, we define contractions at each level:

$$r_{\text{tactic}} \geq r_{\text{technique}} \geq r_{\text{sub-technique}}$$

Following Hutchinson's Theorem, this IFS converges to a unique attractor $K_C$, our "campaign snowflake":

$$K_C = \bigcup_{i=1}^n f_i(K_C)$$

The resulting fractal object $K_C$ serves as a unique geometric signature for campaign $C$.

## 3.3 Hutchinson Operator and Convergence

The Hutchinson operator $\mathcal{H}$ is defined as:

$$\mathcal{H}(S) = \bigcup_{i=1}^n f_i(S)$$

for any set $S \subset \mathbb{R}^2$. Hutchinson's Theorem guarantees that repeated application of this operator converges to the unique attractor:

$$K_C = \lim_{m \to \infty} \mathcal{H}^m(S_0)$$

for any initial non-empty compact set $S_0$.

## 3.4 Fractal Dimension and Campaign Complexity

The box-counting dimension of the attractor $K_C$ provides a quantitative measure of campaign complexity:

$$\dim_{\text{box}}(K_C) = \lim_{\epsilon \to 0} \frac{\log N(\epsilon)}{\log(1/\epsilon)}$$

where $N(\epsilon)$ is the minimum number of squares of side length $\epsilon$ needed to cover $K_C$. For self-similar fractals generated by our IFS with uniform contraction factors, this dimension can be approximated by the similarity dimension:

$$\dim_{\text{sim}}(K_C) = \frac{\log(n)}{\log(1/r)}$$

where $n$ is the number of contractions and $r$ is the contraction factor (assuming all factors are equal).

When implementing strict hierarchical contraction factors $(r_{\text{tactic}} > r_{\text{technique}} > r_{\text{sub-technique}})$ the similarity dimension is instead determined by the unique value of $(d)$ that satisfies the implicit equation:

$$\sum_i r_i^d = 1$$

This equation can be expanded for our three-level hierarchy as:

$$[\sum_{i=1}^{n_{\text{tactic}}} (r_{\text{tactic}})^d + \sum_{j=1}^{n_{\text{technique}}} (r_{\text{technique}})^d + \sum_{k=1}^{n_{\text{sub-technique}}} (r_{\text{sub-technique}})^d = 1]$$

Which simplifies to:

$$[n_{\text{tactic}} \cdot (r_{\text{tactic}})^d + n_{\text{technique}} \cdot (r_{\text{technique}})^d + n_{\text{sub-technique}} \cdot (r_{\text{sub-technique}})^d = 1]$$

Where $(n_{\text{tactic}})$, $(n_{\text{technique}})$, and $(n_{\text{sub-technique}})$ represent the count of elements at each level.

### 3.5 Comparison Metrics: Hausdorff

For comparing campaigns, we employ the Hausdorff distance:

$$d_H(K_{C_1}, K_{C_2}) = \max \left\{ \sup_{x \in K_{C_1}} \inf_{y \in K_{C_2}} d(x,y), \sup_{y \in K_{C_2}} \inf_{x \in K_{C_1}} d(x,y) \right\}$$

## 4 Methodology

### 4.1 Snowflake Generation Algorithm

The transformation from hierarchical taxonomy to fractal snowflake follows a deterministic process:
1. Map campaign tactics to central nodes with angular separation $\theta = 2\pi/n$ for $n$ tactics
2. For each tactic node, generate technique branches with specified contraction factor $r_{\text{tech}}$
3. For each technique branch, generate sub-technique nodes with contraction factor $r_{\text{subtech}}$

The algorithm implements Hutchinson's operator $\mathcal{H}$ iteratively:

$$X_{n+1} = \mathcal{H}(X_n) = \bigcup_{i=1}^{m} f_i(X_n)$$

Where $f_i$ are the contractions corresponding to taxonomy elements, producing the attractor $K_C$ as $n \to \infty$.

## 4.2 Snowflake Generation Algorithm

Campaign similarity is assessed through a three-stage process:
1. Apply the Hutchinson operator to produce attractors $K_{C1}$ and $K_{C2}$
2. Compute the Hausdorff distance $d_H(K_{C1}, K_{C2})$
3. Convert to similarity score via $\text{Sim}(C_1, C_2) = e^{-\lambda d_H(K_{C_1}, K_{C_2})}$

For campaign subset detection (e.g., $C_3 \subset C_1$), we examine both the Hausdorff distance (approaching zero) and the dimensional relationship:

$$d_H(K_{C_3}, K_{C_1}) \approx 0 \quad \text{and} \quad \dim(K_{C_3}) < \dim(K_{C_1})$$

# 5 Case Study

To demonstrate the versatility and robustness of our fractal-based framework for hierarchical attack taxonomy modeling, we present four case studies spanning both cyber and disinformation campaigns. These examples highlight the ability of our method to model campaigns at different levels of granularity, detect subset relationships, and compare structurally similar yet distinct operations. The selected campaigns include both Advanced Persistent Threat (APT) operations and disinformation campaigns, illustrating the domain-agnostic nature of our approach.

## 5.1 APT29: Modeling an APT Group-Level Campaign

APT29, also known as "Cozy Bear," is a sophisticated Russian state-sponsored threat actor group. Its operations typically focus on cyber-espionage targeting government entities, think tanks, and private sector organizations. At the APT group level, we model APT29's general tactics and techniques based on MITRE ATT&CK, but we choose to not be exhaustive as it is not useful for the sake of the mathematical demonstration.

**Table 1.** APT29 Group-level MITRE ATT&CK mapping (simplified)

| Tactic | Technique | Sub-technique |
|---|---|---|
| Initial Access | Phishing | Spearphishing Attachments |
| Initial Access | Supply Chain Compromise | Compromise Software Supply Chain |
| Execution | PowerShell | N/A |
| Persistence | Registry Run Keys/Startup Folder | N/A |
| Persistence | Abuse Elevation Control Mechanism | Bypass User Account Control |
| Defense Evasion | Process Injection | N/A |
| Command and Control | Application Layer Protocol | Web Protocols |

**Fractal representation**
- Center: APT29 aggregated campaigns tactics, techniques and procedures.
- First Circle: 5 tactics
- Second Circle: 7 techniques branching from tactics.
- Third Circle: 3 sparse sub-techniques.

**Fractal dimension**

$$\dim(K_{APT29}) = \frac{\log(n)}{\log(1/r)} = \frac{\log(15)}{\log(3)} = 2.46$$

where $n$=5 tactics + 7 techniques + 3 sub-techniques = 15 total branches, and assumed $r=\frac{1}{3}$ for simplicity.
A python code for implicit equation is available at Appendix 3.

## 5.2 APT29 SolarWinds Campaign: Subset Detection

The SolarWinds campaign, attributed to APT29, represents a highly targeted subset of the group's broader operational capabilities. This campaign exploited the SolarWinds Orion platform to compromise supply chains and infiltrate U.S. government agencies. At the APT specific campaign level, we model APT29's general tactics and techniques based on MITRE ATT&CK, but we choose to not be exhaustive as it is not useful for the sake of the mathematical demonstration.

**Table 2.** APT29's Solarwinds MITRE ATT&CK mapping (simplified)

| Tactic | Technique | Sub-technique |
|---|---|---|
| Initial Access | Supply Chain Compromise | Compromise Software Supply Chain |
| Persistence | Abuse Elevation Control Mechanism | Bypass User Account Control |
| Defense Evasion | Masquerading | Match Legitimate Name or Location |

## 5.3 "Matriochka" Disinformation Campaign

The "Matriochka" campaign, analyzed by VIGINUM (SGDSN, French Government), represents a complex disinformation operation attributed to Russian actors. Its goal was to manipulate public opinion by creating layers of false narratives ("matryoshka dolls") that concealed their origins. (See Appendix n°1 for DISARM mapping)

## 5.4 APT28 DNC Hack (influence campaign)

APT28 (Fancy Bear), another Russian state-sponsored group, conducted a disinformation operation targeting the Democratic National Committee (DNC) during the U.S. presidential election in 2016. This operation combined cyber intrusion with narrative manipulation to influence public opinion. (See Appendix n°2 for DISARM mapping)

## 5.5 Results

These case studies demonstrate that our fractal-based framework effectively models hierarchical attack structures across domains, detects subset relationships (e.g., SolarWinds ⊂ APT29), and quantifies campaign complexity through fractal dimensions (code available at Appendix 3).

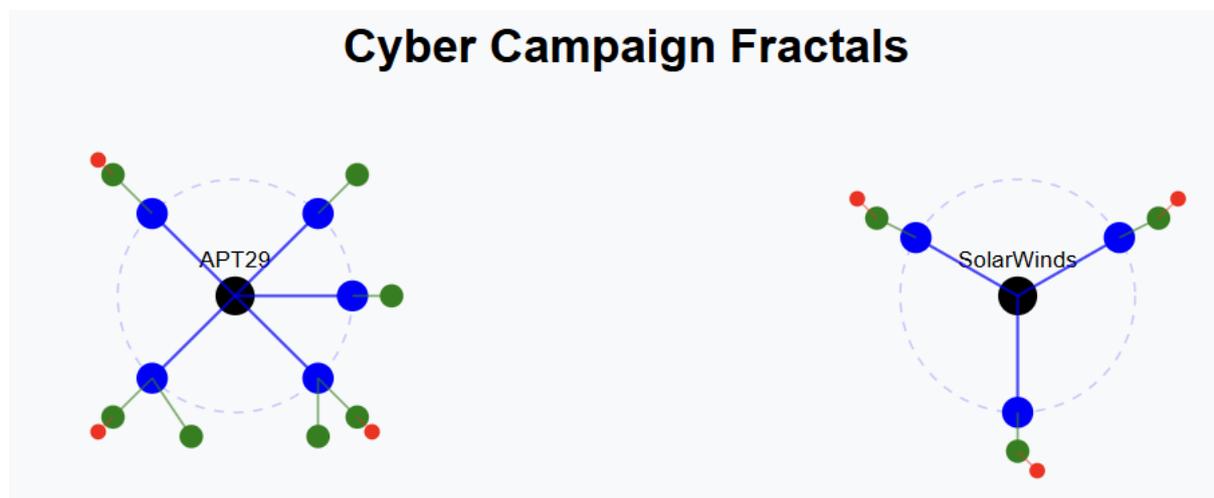

Figure 1 – Exemple of a fractal representation of the APT29 and the SolarWinds case studies

**Dimension analysis**
The hierarchical fractal dimensions using hierarchical contraction factors

$$(r_{\text{tactic}} = 0{,}6 \,;\, r_{\text{technique}} = 0{,}4 \,;\, r_{\text{sub-technique}} = 0{,}2)$$

are consistently higher than those calculated with the simplified formula.

Table 5. Comparative Analysis of Fractal Dimensions

| Fractal | Structure | Dimension (uniform contractor) | Dimension (hierarchical contractor) |
|---|---|---|---|
| APT29 | 5 tactics, 7 techniques, 3 subtechniques | 2.46 | 2.74 |
| SolwarWinds | 3 tactics, 3 techniques, 3 subtechniques | 2.00 | 2.51 |
| Matriochka | 6 tactics, 19 techniques, 19 subtechniques | 3.44 | 4.52 |
| DNCHack | 3 tactics, 6 techniques, 6 subtechniques | 2.46 | 3.17 |

This indicates that the simplified approach underestimates campaign complexity when properly accounting for the hierarchical nature of security taxonomies, but:
1. Relative rankings remain consistent: Matriochka is still the most complex campaign, followed by DNC Hack and APT29, and finally SolarWinds.
2. Hierarchical importance is properly represented: The corrected approach gives appropriate weight to the different taxonomic levels.
3. Subset relationship is preserved: SolarWinds still has a lower dimension than APT29, confirming its subset status.

Despite both APT29 and DNC Hack having the same total number of elements (15) in our use case, their different distribution across taxonomic levels results in distinct fractal dimensions when using hierarchical contraction factors.

This mathematical outcome actually aligns with an important aspect of threat analysis:
- Higher level of detail at the sub-technique level: DNC Hack uses fewer tactics but implements them with more specific sub-techniques, representing a more detailed implementation of a narrower tactical approach.
- Tactical breadth vs. technical depth: APT29 employs more tactics (broader approach) but with less elaboration at the sub-technique level, while DNC Hack demonstrates greater technical depth within a more focused tactical scope.

This analysis demonstrates that while the simplified formula provides a useful approximation, the implicit equation approach better captures the hierarchical nature of security taxonomies.

**Distance**
Using the hierarchical contraction factors with r_tactic=0.6, r_tech=0.4, and r_subtech=0.2, we obtain:

Table 6. Hausdorff Distance Calculation for Cyber Campaign Fractals

| Campaign pair | Hausdorff distance |
|---|---|
| APT29 - SolwarWinds | 0.384 |
| Matriochka - DNCHack | 0.640 |

The Hausdorff distance between APT29 and SolarWinds (0.384) is significantly smaller than between Matriochka and DNC Hack (0.640), which aligns with the paper's settings that SolarWinds campaign is a subset of APT29's broader capabilities.
A python code for implicit equation is available at Appendix 4.

**Sensitivity analysis**

To understand how contraction factors affect the Hausdorff distance calculation, we performed additional computations with varying parameters:

Table 7. Impact of Contraction Factor on Hausdorff Distance Calculation for Cyber Campaign Fractals

| Contraction factors | APT29-SolarWinds | Matriochka-DNCHack |
|---|---|---|
| r_t=0.5, r_te=0.3, r_s=0.1 | 0.3200 | 0.5333 |
| r_t=0.6, r_te=0.4, r_s=0.2 | 0.3840 | 0.6400 |
| r_t=0.7, r_te=0.5, r_s=0.3 | 0.4480 | 0.7467 |

While the absolute values differ substantially, the relative ordering remains consistent: the APT29-SolarWinds relationship consistently demonstrates greater similarity than Matriochka-DNCHack across all parameter choices.

# 6 Discussion

This work demonstrates the versatility of a fractal-based framework for modeling and comparing hierarchical threat campaigns across domains, such as computer network operations and disinformation.

**Using fractal with hierarchical contraction factors**

While our hierarchical fractal representation uses contraction factors where $r_{tactic} > r_{technique} > r_{subtechnique}$ to emphasize taxonomic structure, this ordering produces an interesting effect on dimension calculations. Campaigns with more sub-techniques relative to tactics (like DNC Hack) yield higher dimensions than campaigns with more tactics but fewer sub-techniques (like APT29), even when total element counts are identical. This mathematically captures how technical depth at lower taxonomy levels can contribute more to overall campaign complexity than tactical breadth—an insight aligned with threat analysis practice where sophisticated actors demonstrate more their capabilities at the technical implementations level rather than employing a wide range of tactics, emphasizing the discriminative nature of techniques over tactics.

**On the choice of contraction factors**

The sensitivity analysis demonstrates that the choice of contraction factors, while preserving insigths, significantly impacts the absolute values of Hausdorff distances. This has important methodological implications:
1. Practitioners should calibrate their interpretation of Hausdorff distances based on the specific contraction factors used, rather than relying on absolute thresholds.
2. Threat analysts should focus on relative relationships between campaign pairs rather than absolute distance values.
3. For operational use, the security community would benefit from standardized contraction factors to enable consistent cross-organizational comparisons.

The sensitivity analysis reveals that while the absolute values of Hausdorff distances are highly dependent on parameter choices, the fundamental threat analysis insights remain robust. The fractal framework continues to provide valuable comparative information about campaign relationships, supporting attribution decisions and complexity assessments despite parameter variation.

**Integrated Cyber-Disinformation Campaigns**

The DNC Hack by APT28 exemplifies a hybrid campaign combining cyber intrusion (e.g., data exfiltration) with disinformation (e.g., leak amplification). Our framework can seamlessly integrate such multi-domain operations by introducing an innermost concentric layer branching into domain-specific frameworks like MITRE ATT&CK (cyber) and DISARM (disinformation). Subsequent layers represent their respective tactics, techniques, and sub-techniques. This unified representation enables cross-domain analysis (cyber operations, information warfare, physical security) while preserving the hierarchical integrity of each framework.

**Comparison to Previous Work**

In your previous work on APT behavioral signatures [20], the focus was on paired techniques—capturing interrelations between techniques within a campaign. While this approach excels in identifying operational patterns, it lacks the hierarchical perspective provided by our fractal model. Conversely, our framework emphasizes the branching structure of campaigns but does not explicitly capture interrelations within a single concentric layer (e.g., dependencies between techniques). This highlights complementary strengths:

- Fractal Model: Ideal for quantifying complexity and detecting structural overlaps or subsets when we have a broad coverage of the mapped behavior.
- Paired Techniques: Better suited for analyzing intra-layer relationships and operational dependencies when information is more limited about the observed campaign's behavior.

Future work could combine these approaches to create a hybrid model that captures both hierarchical branching and interrelations within layers.

**Limitations**

While effective for hierarchical modeling, this approach does not capture dependencies within a single layer (e.g., paired techniques in ATT&CK). This limits its ability to analyze operational sequences.

The framework relies on precise mappings to established taxonomies (e.g., DISARM, ATT&CK). Errors or omissions can distort results.

Calculating Hausdorff distances becomes resource-intensive for large datasets.

The sensivity analysis highlight the importance of careful parameter selection and standardization for operational applications in threat intelligence, that is not covered in this research.

# 7  Conclusion

We have presented a novel approach to threat campaign analysis through fractal geometry, establishing a mathematical foundation for comparing hierarchical attack structures across both cyber and information security domains. By conceptualizing taxonomic elements from frameworks like MITRE ATT&CK and DISARM as snowflake-like fractal patterns, we enable both computational precision and intuitive visualization of complex threat landscapes.

The application of Hutchinson's Theorem provides a formal mechanism for generating unique campaign signatures regardless of domain, while Hausdorff distance offers a principled metric for assessing structural relationships between campaigns. Our case studies demonstrate the approach's ability to detect campaign subsets (e.g., SolarWinds within APT29), measure similarity between independent operations, and quantify complexity across cyber operations and disinformation campaigns alike.

This work establishes a new bridge between security taxonomy research and computational geometry, offering analysts both mathematical rigor and visual intuition for addressing the growing complexity of adversarial campaigns. The domain-agnostic nature of our framework makes it particularly valuable for analyzing hybrid threats that span multiple domains, such as the APT28 DNC Hack which combined cyber intrusion with disinformation tactics.

As adversarial tactics continue to evolve across both cyber and information domains, geometric approaches may prove increasingly valuable in recognizing patterns across seemingly disparate operations, enhancing attribution capabilities, and ultimately improving defensive postures against sophisticated threats.

**Appendix n°1 - "Matriochka" Disinformation Campaign DISARM Mapping**

| Tactic | Technique | Sub-technique |
|---|---|---|
| Narrative Development | Create Fake Personas | Use Fake Academic Credentials |
| Narrative Development | Develop Divisive Content | Emotional Content Targeting |
| Narrative Development | Amplify Existing Narratives | Exploit Polarizing Topics |
| Narrative Development | Fabricate Narratives | Create False News Articles |
| Platform Manipulation | Amplify Messages | Cross-Platform Coordination |
| Platform Manipulation | Hijack Trending Topics | Use Hashtag Hijacking |
| Platform Manipulation | Exploit Platform Algorithms | Optimize Content for Algorithmic Boost |
| Platform Manipulation | Spread Malicious Links | Use Shortened URLs |
| Audience Targeting | Segment Audiences | Target Based on Demographics |
| Audience Targeting | Deploy Microtargeting | Use Behavioral Data |
| Audience Targeting | Exploit Cognitive Biases | Appeal to Emotions |
| Establish Legitimacy | Create Fake Experts | Use Fake Credentials |
| Establish Legitimacy | Develop Front Organizations | Use Fake NGOs |
| Establish Legitimacy | Hijack Trusted Sources | Impersonate Journalists |
| Discredit Opponents | Spread Negative Narratives | Fabricate Scandals |
| Discredit Opponents | Create False Associations | Link Opponents to Extremist Groups |
| Discredit Opponents | Amplify Negative Content | Spread Edited Photos/Videos |
| Platform Manipulation | Exploit Bots | Deploy Automated Accounts |
| Platform Manipulation | Exploit Troll Farms | Coordinate Troll Activity |

(truncated for simplification purpose, limiting to 44 branches)

**Appendix n°2 - APT28 DNC Hack Campaign DISARM Mapping**

| Tactic | Technique | Sub-technique |
|---|---|---|
| Narrative Development | Leak Amplification | Amplify Leaked Documents |
| Narrative Development | Develop Divisive Narratives | Frame Opponents as Corrupt |
| Narrative Development | Amplify Existing Narratives | Exploit Polarizing Topics |
| Platform Manipulation | Amplify Messages | Cross-Platform Coordination |
| Platform Manipulation | Exploit Platform Algorithms | Optimize Content for Algorithmic Boost |
| Audience Targeting | Segment Audiences | Target Based on Political Affiliation |

**Appendix n°3 – Python code used to compute dimensions in the case studies**

```python
import math

def calculate_fractal_dimension(tactics_count, techniques_count, subtechniques_count,
                                r_tactic=0.6, r_tech=0.4, r_subtech=0.2):
    """Calculate the corrected fractal dimension for a hierarchical taxonomy."""
    # Define contraction factors for each level
    tactic_scale = r_tactic
    technique_scale = r_tactic * r_tech  # Compound scaling
    subtechnique_scale = r_tactic * r_tech * r_subtech  # Compound scaling

    # Define the implicit equation ∑_i r_i^d = 1
    def f(d):
        return (tactics_count * (tactic_scale ** d) +
                techniques_count * (technique_scale ** d) +
                subtechniques_count * (subtechnique_scale ** d) - 1)

    # Bisection method to solve for d where f(d) = 0
    low, high = 0.1, 10.0

    for _ in range(100):  # Maximum 100 iterations
        mid = (low + high) / 2
        val = f(mid)

        # If the value is sufficiently close to zero, break
        if abs(val) < 1e-12:
            break

        if f(low) * val < 0:
            high = mid
        else:
            low = mid

    return mid

# Case studies from the paper
case_studies = [
    {"name": "APT29", "tactics": 5, "techniques": 7, "subtechniques": 3},
    {"name": "SolarWinds", "tactics": 3, "techniques": 3, "subtechniques": 3},
    {"name": "Matriochka", "tactics": 6, "techniques": 19, "subtechniques": 19},
    {"name": "DNC Hack", "tactics": 3, "techniques": 6, "subtechniques": 6}
]

# Calculate dimensions for all case studies
for case in case_studies:
    # Calculate original dimension using simplified formula
    n = case["tactics"] + case["techniques"] + case["subtechniques"]
    r = 1/3  # As used in the paper
    original_dim = math.log(n) / math.log(1/r)

    # Calculate corrected dimension using hierarchical factors
    corrected_dim = calculate_fractal_dimension(
        case["tactics"],
        case["techniques"],
        case["subtechniques"]
    )

    print(f"{case['name']} - Original: {original_dim:.2f}, Corrected: {corrected_dim:.2f}")
```

**Appendix n°4 – Python code used to compute Hausdorff in the case studies**

```
import numpy as np
from scipy.spatial.distance import directed_hausdorff
import matplotlib.pyplot as plt

def generate_fractal_points(tactics, techniques, subtechniques,
                            r_tactic=0.6, r_tech=0.4, r_subtech=0.2):
    """
    Generate points representing a fractal snowflake for a campaign
    with hierarchical contraction factors
    """
    points = []

    # Center point
    center = np.array([0, 0])
    points.append(center)

    # Generate tactic points in a circle around center
    tactic_points = []
    tactic_angles = np.linspace(0, 2*np.pi, tactics, endpoint=False)

    for angle in tactic_angles:
        x = r_tactic * np.cos(angle)
        y = r_tactic * np.sin(angle)
        tactic_point = np.array([x, y])
        tactic_points.append(tactic_point)
        points.append(tactic_point)

    # Distribute techniques among tactics
    techniques_per_tactic = techniques // tactics
    remaining_techniques = techniques % tactics

    technique_points = []

    for i, tactic_point in enumerate(tactic_points):
        # Calculate how many techniques branch from this tactic
        n_techniques = techniques_per_tactic
        if i < remaining_techniques:
            n_techniques += 1

        if n_techniques == 0:
            continue

        # Generate technique points in a circle around the tactic
        tech_angles = np.linspace(0, 2*np.pi, n_techniques, endpoint=False)

        for angle in tech_angles:
            # Apply compound scaling: tactic scale * technique scale
            tech_radius = r_tactic * r_tech
            x = tactic_point[0] + tech_radius * np.cos(angle)
            y = tactic_point[1] + tech_radius * np.sin(angle)
            technique_point = np.array([x, y])
            technique_points.append(technique_point)
            points.append(technique_point)

    # Distribute subtechniques among techniques
    subtechniques_per_technique = subtechniques // max(1, len(technique_points))
    remaining_subtechniques = subtechniques % max(1, len(technique_points))

    for i, technique_point in enumerate(technique_points):
        # Calculate how many subtechniques branch from this technique
        n_subtechniques = subtechniques_per_technique
        if i < remaining_subtechniques:
            n_subtechniques += 1

        if n_subtechniques == 0:
            continue

        # Generate subtechnique points in a circle around the technique
        subtech_angles = np.linspace(0, 2*np.pi, n_subtechniques, endpoint=False)

        for angle in subtech_angles:
            # Apply compound scaling: tactic scale * technique scale * subtechnique scale
            subtech_radius = r_tactic * r_tech * r_subtech
            x = technique_point[0] + subtech_radius * np.cos(angle)
```

```python
            y = technique_point[1] + subtech_radius * np.sin(angle)
            subtechnique_point = np.array([x, y])
            points.append(subtechnique_point)

    return np.array(points)

def calculate_hausdorff_distance(campaign1, campaign2, r_tactic=0.6, r_tech=0.4, r_subtech=0.2):
    """
    Calculate the Hausdorff distance between two campaign fractals
    """
    # Generate points for both campaigns
    points1 = generate_fractal_points(*campaign1, r_tactic, r_tech, r_subtech)
    points2 = generate_fractal_points(*campaign2, r_tactic, r_tech, r_subtech)

    # Compute the directed Hausdorff distances
    forward_hausdorff, _, _ = directed_hausdorff(points1, points2)
    backward_hausdorff, _, _ = directed_hausdorff(points2, points1)

    # Hausdorff distance is the maximum of the two directed distances
    hausdorff_distance = max(forward_hausdorff, backward_hausdorff)

    return hausdorff_distance

# Define the four campaigns from the case studies
apt29 = (5, 7, 3)   # 5 tactics, 7 techniques, 3 subtechniques
solarwinds = (3, 3, 3)  # 3 tactics, 3 techniques, 3 subtechniques
matriochka = (6, 19, 19)  # 6 tactics, 19 techniques, 19 subtechniques
dnc_hack = (3, 6, 6)  # 3 tactics, 6 techniques, 6 subtechniques

# Calculate Hausdorff distances with hierarchical contraction factors
r_tactic = 0.6
r_tech = 0.4
r_subtech = 0.2

# Compare APT29 and SolarWinds
hausdorff_apt29_solarwinds = calculate_hausdorff_distance(
    apt29, solarwinds, r_tactic, r_tech, r_subtech)

# Compare Matriochka and DNC Hack
hausdorff_matriochka_dnchack = calculate_hausdorff_distance(
    matriochka, dnc_hack, r_tactic, r_tech, r_subtech)

print(f"Hausdorff distance between APT29 and SolarWinds: {hausdorff_apt29_solarwinds:.6f}")
print(f"Hausdorff distance between Matriochka and DNC Hack: {hausdorff_matriochka_dnchack:.6f}")
```